\documentclass[aps,showpacs,prl,toolkits,twocolumn]{revtex4}
\usepackage{epsfig,amssymb,amsfonts,amsmath,graphicx}

\makeatletter
\renewcommand{\@makefntext}[1]{\parindent=1em\noindent\hbox to 1.8em
{\hss$^{\@thefnmark}$}#1}
\renewcommand{\@footnotemark}{\hbox{\mathsurround=0pt$^{\@thefnmark}$}}

\makeatother

\begin{document}
\title{Chirally symmetric but confining dense and cold matter.}
\author{ L. Ya. Glozman and R. F. Wagenbrunn}
\affiliation{Institute for
 Physics, Theoretical Physics branch, University of Graz, Universit\"atsplatz 5,
A-8010 Graz, Austria}

\newcommand{\be}{\begin{equation}}
\newcommand{\bea}{\begin{eqnarray}}
\newcommand{\ee}{\end{equation}}
\newcommand{\eea}{\end{eqnarray}}
\newcommand{\ds}{\displaystyle}
\newcommand{\low}[1]{\raisebox{-1mm}{$#1$}}
\newcommand{\loww}[1]{\raisebox{-1.5mm}{$#1$}}
\newcommand{\lmn}{\mathop{\sim}\limits_{n\gg 1}}
\newcommand{\vpint}{\int\makebox[0mm][r]{\bf --\hspace*{0.13cm}}}
\newcommand{\too}{\mathop{\to}\limits_{N_C\to\infty}}
\newcommand{\vp}{\varphi}
\newcommand{\vx}{{\vec x}}
\newcommand{\vy}{{\vec y}}
\newcommand{\vz}{{\vec z}}
\newcommand{\vk}{{\vec k}}
\newcommand{\vq}{{\vec q}}
\newcommand{\vpp}{{\vec p}}
\newcommand{\vn}{{\vec n}}
\newcommand{\vg}{{\vec \gamma}}

\begin{abstract}
The folklore tradition about the QCD phase diagram is that at the
chiral restoration phase transition at finite density hadrons are
deconfined and there appears the quark matter. 
 We address this question within
the only known exactly solvable confining and chirally symmetric model.
It is postulated within this model that there exists linear Coulomb-like
confining interaction. The chiral symmetry breaking and the quark
Green function are obtained from the Schwinger-Dyson (gap) equation while
the color-singlet meson spectrum results from the Bethe-Salpeter
equation.  We solve this model at T=0
and finite chemical potential $\mu$ and obtain a clear chiral restoration
phase transition at the critical value $\mu_{cr}$. Below this value
 the spectrum 
is similar to the previously obtained one at $\mu = 0$. At $\mu > \mu_{cr}$
the quarks are still confined and the physical spectrum consists of bound
states which are arranged into a complete set of exact chiral multiplets.
This explicitly demonstrates that  a chirally symmetric matter
consisting of confined but chirally symmetric hadrons
at finite chemical potential is also possible in QCD.  If so, there must 
be nontrivial implications for astrophysics.
\end{abstract}
\pacs{11.30.Rd, 25.75.Nq, 12.38.Aw}

\maketitle

\section{Introduction}

The structure of the QCD phase diagram is one of the keys for our 
understanding of QCD, strongly interacting matter and has direct astrophysical
implications. It is a subject of a significant effort within the
heavy ion collision programs at different labs. 
We know that in the vacuum as well as at low temperature and
density QCD is realised in the confining 
hadronic phase with broken chiral symmetry. At  high temperature and 
low density the system
is deconfined and chiral symmetry is restored.
Above the critical temperature
one experimentally observes a strongly interacting plasma with dynamics
that is not yet well understood.
Only at the very
high temperatures one possibly approaches a perturbative regime of
the weakly interacting quark-gluon plasma.  Just in the opposite limit of the 
asymptotically large densities and small temperature one believes that
the perturbative gluonic interaction is also adequate and consequently one has, 
depending
on the number of active flavors
the color superconducting phase (the color-flavor locking phase with $N_f = N_c$). 

There is a folklore
tradition that upon increasing temperature and density from zero
the deconfining and chiral restoration transitions 
coincide
and beyond the semi-circle in the $T-\mu$ plane one obtains a deconfining
and chirally restored matter. From
the theoretical side we know from the 't Hooft anomaly matching conditions
\cite{hooft} that at zero temperature and density in the confining phase
chiral symmetry must be necessarily broken in the vacuum. 
The converse is not required, of course.
Whether the chiral restoration and deconfinement transitions coincide
at finite temperature and zero chemical
potential can be answered on the lattice. It was believed that the same
temperature triggers both transitions \cite{karsch}, though recently there
have appeared indications that it could be not so \cite{fodor}. 
Unfortunately lattice cannot help us at finite chemical potential
due to the notorious sign problem. Hence  one has to rely
on  models  of QCD. Typically  models that give us 
information about the
phase structure are of the Nambu and Jona-Lasinio type, for an overview
and references see \cite{stephanov}. An obvious drawback of all these
models is that they are not confining. Then there is no basis to conclude
that above the chiral restoration point at finite chemical potential
one obtains a chirally symmetric and deconfining quark matter.

Recently McLerran and Pisarski presented qualitative large $N_c$ arguments
showing that at a reasonably large chemical potential and low temperature
there might exist a confining but chirally symmetric phase  \cite{pisarski}.
  This 
suggestion would dramatically
change the existing paradigm 
and requires urgent efforts to verify whether
it could be indeed the case. This suggestion is in conflict with
the naive intuition that once the hadrons are confined chiral symmetry
should be necessarily broken. There are no such limitations from
the 't Hooft anomaly matching conditions at finite density, however,
and such an intuition is in fact  an artefact of  existing simple models
of chiral symmetry breaking and confinement \cite{BAG,CASHER}.

 Clearly we cannot solve QCD and answer this
question from  first principles. What can be done is to verify 
this conjecture within
a solvable  manifestly confining and chirally symmetric model.  
Such a model could provide  insight into
physics that would drive such an unexpected situation. If confirmed 
within such a model, this scenario 
could 
also be realised in QCD and further theoretical and
experimental efforts to clarify this important question would be required.

There exists only one known manifestly chirally-symmetric and confining
model in four dimensions that is solvable \cite{Orsay}, sometimes
called  Generalized Nambu and Jona-Lasinio model (GNJL). This model
can be considered as a generalization of the 1+1 dimensional
't Hooft  model, that is QCD in the large $N_c$ limit \cite{HOOFT}.
Once the gauge is properly chosen in 1+1 dimensions 
the Coulomb interaction becomes a linear confining
potential.  Then  this
potential properly represents gluonic degrees of freedom in 1+1
dimensions. In four dimensions
there are different gluonic interactions and it is hopeless to solve
even large $N_c$ QCD with full gluodynamics.
It is postulated within the GNJL model that there
exists a linear confining potential of the Coulomb type in four 
dimensions \footnote{Actually there is a number of works that derive
such a linear rising confining potential of the Coulomb type in the Coulomb
gauge QCD \cite{linear} and that connect this potential with the
Gribov-Zwanziger scenario for confinement in the Coulomb gauge \cite{GZ}.}.
This model represents
a simplification of  large $N_c$ QCD.  The model is exactly solvable. 
The chiral symmetry breaking 
and the properties of the Goldstone bosons can
be obtained from the solution of the Schwinger-Dyson and Bethe-Salpeter
equations  \cite{Adler:1984ri,Alkofer:1988tc,BR,BN,COT,W}. The complete
spectrum of $\bar q q$ mesons has been obtained in refs. \cite{WG1,WG2},
which exhibits restoration of chiral symmetry in excited hadrons, for
a review see \cite{GPR}. 

It is instructive to review the chiral symmetry and confining properties
of quarks in this model, which are similar to those in the 't Hooft model.
Upon solving the Schwinger-Dyson equation for the quark Green functions
one obtains chiral symmetry breaking in the vacuum, i.e. a nonzero quark
condensate and a dynamical mass $M(q)$ for the quarks, that is strongly momentum
dependent and vanishes at large momenta. This single quark Green function
is infrared-divergent, which means that the single quark cannot be observed
and hence is confined. The infrared divergence is canceled exactly in the
Bethe-Salpeter equation for the
observable color-singlet hadrons and the physical spectrum consists only of
 color-singlet hadrons. This model allows for a solution at finite
chemical potential and therefore can be used as a tool to check whether it is
possible or not to have  confining but chirally symmetric matter at
finite chemical potential. Below we solve it and obtain a complete
spectrum of mesons at zero temperature and
finite quark chemical potential. We demonstrate
that below the critical chemical potential, $\mu_{cr}$, the spectrum
is similar to the one previously obtained  at zero chemical potential
\cite{WG1,WG2}. At  $\mu =\mu_{cr}$ the chiral restoration phase
transition happens. At $\mu > \mu_{cr}$  hadrons are still confined
and the spectrum represents a complete set of chiral multiplets. This
 explicitly demonstrates that a chirally symmetric but confining phase
exists within this model and opens a perspective to study such a possibility in
QCD.

\section {Chiral symmetry breaking and Bethe-Salpeter equation for mesons in
 vacuum}

Throughout this paper we work in the chiral limit. The two flavor 
 version of the model is considered. The global chiral symmetry
of the model is $U(2)_L \times U(2)_R$, because in the large $N_c$
world the axial anomaly is absent.
The model is described
in great detail in references \cite{WG2,GPR} so here we give only an
overview of this model in  vacuum. 
It is postulated within the model that there is only a linear instantaneous
Lorentz-vector interquark potential that has a Lorentz
structure of the Coulomb
potential, i.e., it is a density-density interaction:

\begin{equation} 
K^{ab}_{\mu\nu}(\vec{x}-\vec{y})=g_{\mu 0}g_{\nu 0}
\delta^{ab} V (|\vec{x}-\vec{y}|); ~~~~~
\frac{\lambda^a \lambda^a}{4}V(r) = \sigma r,
\label{KK}
\end{equation}

\noindent
where $a,b$ are color indices. The Fourier transform of this
potential and any loop integral are infrared-divergent.
Hence  infrared regularization is required and any physical
observable, such as a hadron mass, must be independent of the infrared
regulator $\mu_{IR}$ in the infrared limit (i.e. when this regulator 
approaches 0, $\mu_{IR} \rightarrow 0$ ).
There are several physically equivalent ways to perform this infrared
regularization. Here we follow Ref. \cite{Alkofer:1988tc} and define the
potential in momentum space as

\begin{equation}
V(p)= \frac{8\pi\sigma}{(p^2 + \mu_{\rm IR}^2)^2}.
\label{FV} 
\end{equation}

\noindent
Then this potential in the configuration space contains the
required $\sigma r$ term,  the infrared-divergent term
$-\sigma /\mu_{IR}$ as well as terms that vanish in the infrared limit.

The Dirac operator for the dressed quark is
\begin{equation}
D(p_0,\vec{p})= i S^{-1}(p_0,\vec{p}) = D_0(p_0,\vec{p})-\Sigma(p_0,\vec{p}),
\label{SAB}
\end{equation}

\noindent
where  $D_0$ is the bare Dirac operator. 
Parametrising the self-energy operator in the form

\begin{equation}
\Sigma(\vec p) =A_p +(\vec{\gamma}\cdot\hat{\vec{p}})[B_p-p],
\label{SE} 
\end{equation}

\noindent
where functions $A_p$ and $B_p$ are yet to be found, the
Schwinger-Dyson equation for the self-energy operator in 
the rainbow approximation,which is valid in the large $N_c$ limit
for the instantaneous interaction, 
 is reduced to the nonlinear gap equation for the chiral angle
 $\varphi_p$,
 
 \begin{equation}
 A_p \cos \varphi_p - B_p \sin \varphi_p = 0,
 \label{gap}
 \end{equation}
 
\noindent
where
 
\begin{eqnarray}
A_p & = & \frac{1}{2}\int\frac{d^3k}{(2\pi)^3}V
(\vec{p}-\vec{k})\sin\vp_k,\quad  
\label{AB1} \\
B_p & = & p+\frac{1}{2}\int \frac{d^3k}{(2\pi)^3}\;(\hat{\vec{p}}
\cdot\hat{\vec{k}})V(\vec{p}-\vec{k})\cos\vp_k. 
\label{AB2} 
\end{eqnarray}  
 
The functions $A_p,B_p,$  i.e. the quark self-energy,
  are divergent in the
 infrared limit, 
 
 \begin{eqnarray}
A_p&=&\frac{\sigma}{2\mu_{\rm IR}}\sin\varphi_p+A^f_p, 
\label{AA}
\end{eqnarray}
 
 \begin{eqnarray}
B_p&=&\frac{\sigma}{2\mu_{\rm IR}}\cos\varphi_p+B^f_p,
\label{BB}
\end{eqnarray}
where $A^f_p$ and $B^f_p$ are infrared-finite functions.
This     implies that the single quark cannot be observed
 and the system is confined. However, the infrared divergence 
 cancels exactly in 
 the gap equation (\ref{gap})
so this equation can be solved directly in the infrared limit\footnote{%
Actually we have solved the gap equation with small 
values of $\mu_{\rm IR}>0$ in the potential for the IR-divergent 
functions $A_p$ and $B_p$ by iteration on a mesh of a finite number 
of points in momentum space. In both integrals the angular integrations 
can be performed analytically. For the remaining numerical calculations 
of the radial integrals special care has been taken with respect to the 
vicinity of $k=p$ where the integrands have sharp peaks.
The results for all IR-finite quantities have then been extrapolated 
numerically
to $\mu_{\rm IR}=0$.}.
The chiral symmetry breaking is signalled by the nonzero quark condensate
and by the dynamical momentum-dependent mass of quarks

\begin{equation}
\langle\bar{q}q\rangle=-\frac{N_C}{\pi^2}\int^{\infty}_0 dp\;p^2\sin\vp_p,
~~~~~~~
M(p) = p \tan \varphi_p.
\label{dyna}
\end{equation} 

\noindent
The dynamical mass is finite at small momenta and vanishes
at large momenta.
Both these quantities were first obtained in ref. \cite{Adler:1984ri}
and repeatedly reconfirmed in all subsequent works on this model
\cite{Alkofer:1988tc,BR,BN,COT,W,WG1,WG2}. 
The numerical value of the quark condensate is 
$\langle\bar{q}q\rangle=(-0.231\sqrt{\sigma})^3$. 
 
The homogeneous Bethe-Salpeter equation for a quark-antiquark
bound state in the rest frame with the instantaneous interaction
is  

\begin{eqnarray}
\chi(m,\vpp)&= &- i\int\frac{d^4q}{(2\pi)^4}V(|\vpp-\vq|)\;
\gamma_0 S(q_0+m/2,\vpp-\vq) \nonumber \\
& \times & \chi(m,\vq)S(q_0-m/2,\vpp-\vq)\gamma_0.
\label{GenericSal}
\end{eqnarray}

\noindent
Here $m$ is the meson mass and $\vec p$ is the relative momentum. 
The Bethe-Salpeter
equation can be solved by means of expansion of the vertex function
$\chi(m,\vpp)$  into a set of all possible independent 
amplitudes consistent with $I,J^{PC}$. Then the Bethe-Salpeter equation
transforms into a system of coupled equations \cite{WG2}. The infrared
divergence cancels exactly in these equations and they can be solved
either in the infrared limit  or for very small values of the infrared
regulator.

In the limit of vanishing dynamical quark mass $M(p) = 0$ 
(i.e., $\varphi_p = 0$), the Bethe-Salpeter equation transforms
into systems of coupled equations that exactly fall into a complete
set of chiral multiplets \cite{WG2}:

\begin{center}
{\bf J~=~0}

\begin{eqnarray}
(1/2,1/2)_a  &  :  &   1,0^{-+} \longleftrightarrow 0,0^{++}  \nonumber \\
(1/2,1/2)_b  &  : &   1,0^{++} \longleftrightarrow 0,0^{-+} ,
\label{sym1}
\end{eqnarray}
\bigskip
{\bf J~=~2k,~~~k=1,2,...}
\begin{eqnarray}
 (0,0)  & :  &   0,J^{--} \longleftrightarrow 0,J^{++}  \nonumber \\
 (1/2,1/2)_a  & : &   1,J^{-+} \longleftrightarrow 0,J^{++}  \nonumber \\
 (1/2,1/2)_b  & : &   1,J^{++} \longleftrightarrow 0,J^{-+}  \nonumber \\
 (0,1) \oplus (1,0)  & :  &   1,J^{++} \longleftrightarrow 1,J^{--} 
\label{sym2}
\end{eqnarray}

\bigskip
{\bf J~=~2k-1,~~~k=1,2,...}

\begin{eqnarray}
 (0,0)  & :  &   0,J^{++} \longleftrightarrow 0,J^{--}  \nonumber \\
 (1/2,1/2)_a  & : &   1,J^{+-} \longleftrightarrow 0,J^{--}  \nonumber \\
 (1/2,1/2)_b  & : &   1,J^{--} \longleftrightarrow 0,J^{+-}  \nonumber \\
 (0,1) \oplus (1,0)  & :  &   1,J^{--} \longleftrightarrow 1,J^{++} 
\label{sym3}
\end{eqnarray}
\end{center}

\noindent
Here $I, J^{PC}$ is a standard notation for isospin $I$, spin $J$ as well as for
both spatial and charge parities $P$ and $C$.
The sign $\longleftrightarrow$ indicates that both given states belong to
the same representation and that the Bethe-Salpeter equations for these
states are {\it identical}.

The complete spectrum of the $\bar q q$ mesons in the vacuum
can be found in refs. 
\cite{WG1,WG2} and exhibits a fast  chiral restoration with increasing
$J$. A reason for this restoration is that with larger $J$ the radial
wave function vanishes at small relative momenta, and hence the
large chiral symmetry breaking dynamical mass $M(p)$ becomes irrelevant
and asymptotically at $J \rightarrow \infty$ the Bethe-Salpeter equation
exactly decouples into a set of equations (\ref{sym2})-(\ref{sym3}).

\section{Inclusion of a finite quark chemical potential at zero temperature} 

\subsection{The gap equation}

Now we are in a position to include into this model a finite quark
chemical potential at zero temperature. 
We can straightforwardly do it only  at zero temperature because to employ
thermodynamics at finite temperature one would need to know a  
 temperature dependence of the confining interaction. In addition
 there is no screening of the confining interaction 
 at zero temperature in the
 large $N_c$ limit since in this limit there are no  vacuum quark
loops and no the particle-hole modifications of the gluonic propagator due
to  Debye screening \cite{pisarski}. 

The effect of a finite quark
chemical potential at zero temperature reduces to the Pauli 
blocking of all occupied levels.
Denoting the Fermi momentum of quarks as $p_f$ 
one has to replace the vacuum density matrix $v(\vec p) v^\dagger(\vec p)$
by the density matrix in the medium with all quark
levels occupied up to the Fermi momentum:

\begin{equation}
\rho(\vec p) = \Theta(p_f - p) u(\vec p) u^\dagger(\vec p)
+ v(\vec p) v^\dagger(\vec p).
\label{den}
\end{equation}
Hence
one has to remove from the integration
both in the Schwinger-Dyson (gap) and Bethe-Salpeter equations all
quark momenta below $p_f$ since they are Pauli-blocked. 
The modified gap equation  at any $p > p_f$
is then the same as in (\ref{gap}) - (\ref{AB2}),
but the integration starts not from $k=0$, but from $k=p_f$.
The points $p < p_f$ are irrelevant for the hadron
which is always on top of the Fermi sea.
Similar, in the expression (\ref{dyna}) for the quark condensate
the lower limit of the integration is now $p_f$, rather than 0.
As in the vacuum, we are in the position to
solve numerically the nonlinear gap equation at arbitrary $p_f$, study the 
chiral restoration phase transition at the critical $p_f^{cr}$,
and, given a dressed quark Green function, solve the 
Bethe-Salpeter equation for the spectrum below and above the critical
Fermi momentum.

\medskip

In Fig. 1 we show  the quark condensate $\langle \bar q q
\rangle$ as a function of  $p_f$.
\begin{figure}
\includegraphics[width=0.88\hsize,clip=]{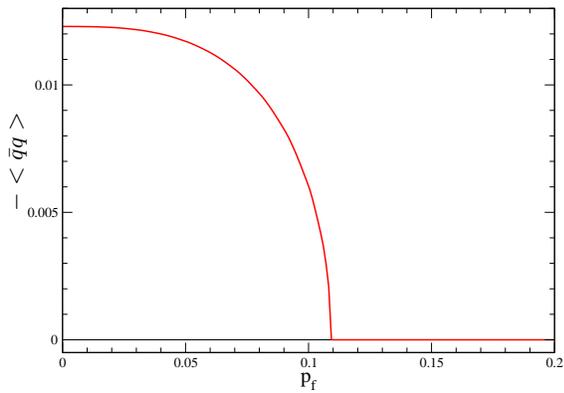}
\caption{Quark condensate in units of $\sigma^{3/2}$
as a function of the Fermi momentum, which is units of $\sqrt \sigma$.}
\end{figure}
One observes an obvious phase transition at the critical Fermi momentum
 $p_f^{cr} = 0.109\sqrt{\sigma}$. 
This is similar to what has been 
observed within
the 't Hooft model \cite{Thies}.
The reason for this chiral
symmetry restoration phase transition is the same as in the standard
Nambu and Jona-Lasinio model \cite{NJL} or in the 't Hooft model \cite{Thies}:
Once a considerable part  of the quark self-energy interaction
is removed by the Pauli blocking, then there
is not enough strength in the gap equation to generate a nontrivial
solution with the broken chiral symmetry in the vacuum.

\begin{figure}
\includegraphics[width=0.8\hsize,clip=]{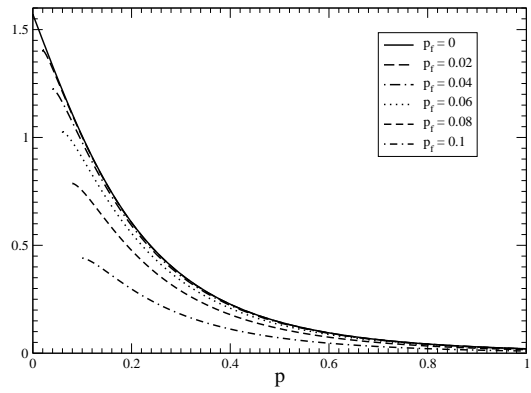}
\caption{Chiral angle $\varphi_p$ as a function of the  momentum $p$
at different fixed values
of the Fermi momentum $p_f$ below the critical value. 
Both momenta are units of $\sqrt \sigma$.}
\end{figure}

In order to see this physics explicitly consider  solutions
of the gap equation for the chiral angle $\varphi_p$ at different fixed
 Fermi momenta $p_f$, that are shown in Fig. 2. One obviously
observes a universal approaching of the nontrivial solution of the
gap equation $\varphi_p \neq 0$ to its trivial value $\varphi_p = 0$
once the Fermi momentum $p_f$ approaches a critical value $p_f^{cr}$.
Note that the infrared point in the gap equation (\ref{gap})-(\ref{AB2})
is $\vec k = \vec p$ (i.e., where the argument of the potential approaches 0).
It directly follows from the self-energy integrals (\ref{AB1})-(\ref{AB2}). 
For any $p > p_f$ 
this point is {\it always} within the integration interval from $k=p_f$ to
$k=\infty$.
However, while the linear confining potential is infrared-singular itself,
the gap equation (\ref{gap}) is not, because the infrared divergence
cancels out exactly, as it follows from  (\ref{AA})-(\ref{BB}). Hence
the contribution to the gap equation comes not only from the
vicinity of the infrared point $\vec p = \vec k$, but actually from the
whole integration interval. Once a critical part of this integration
interval is removed by the Pauli blocking, the nontrivial chiral
symmetry breaking solution of the gap equation abruptly disappears.
A $p_f$ dependence of the dynamical mass $M(p)$ is
  shown in Fig. 3.

Summarizing, we stress  that the symmetry gets restored in our
case not because the potential gets screened. In the large $N_c$
limit it cannot be screened and the string tension $\sigma$ remains
constant at any $p_f$. It gets restored exclusively due to the Pauli 
blocking that reduces a "strength" of the gap equation below the critical
value. At each  fixed value of $\sigma$ there always exists such a critical
reduction of the integration interval due to the Pauli blocking so that
the nontrivial chiral symmetry breaking solution vanishes. This
directly follows from the fact that {\it all} dimensional quantities
in our task are given in units of $\sigma$ and all results are valid
for any $\sigma$, large or small.

Above the critical value $p_f^{cr}$ there is no nontrivial chiral
symmetry breaking solution and the only solution is trivial, $\varphi_p=0$.
This follows from the numerical integration of the highly nonlinear
gap equation. It would be difficult, if impossible, to obtain this
critical Fermi momentum analytically.

Hence the chiral symmetry gets restored, $\varphi_p = 0$, with
vanishing  quark condensate, $\langle \bar q q\rangle = 0$,
and dynamical mass of quarks, $M(q) = 0$, as it follows
from (\ref{dyna}).
At $\varphi_k = 0$  the Lorentz-scalar self-energy of quarks  vanishes
identically,
$A_p=0$, see eq. (6).

 \begin{figure}
\includegraphics[width=0.8\hsize,clip=]{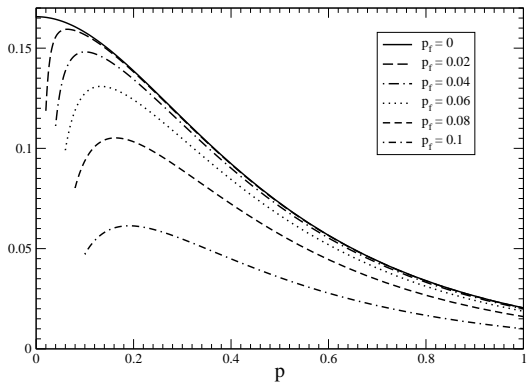}
\caption{Dynamical mass of quarks $M(p)$
as a function of the  momentum $p$ at different fixed values
of the Fermi momentum $p_f$ below the critical value. 
Both momenta are units of $\sqrt \sigma$.}
\end{figure}

What
crucially distinguishes this model from the NJL model is that the quark
is still confined, even in the chirally restored phase. This is because
for any $p > p_f$
the Lorentz spatial-vector self-energy integral $B_p$ 
does not vanish at $\varphi_k = 0$  (see eq. (\ref{AB2})) and
is in fact
infrared-divergent, see (\ref{BB}). 
Hence the single quark 
 is removed from the spectrum at any chemical potential.

\subsection{Meson spectrum}

 This infrared divergence of the single quark Green function
cancels exactly, however, in the color-singlet quark-antiquark system
\cite{WG2} and the bound state mesons are finite and well defined
quantities. Like in the gap equation, the only modification of the
Bethe-Salpeter equation is that the integration in $q$ starts not
from 0, but from $q = p_f$.

The complete spectrum of $\bar q q$ mesons at the Fermi momentum $p_f$
essentially below the critical value, $p_f = 0.05\sqrt{\sigma}$, 
is shown on Fig. 4. This spectrum is similar to the previously 
obtained one in the vacuum
in refs. \cite{WG1,WG2}. Four Goldstone bosons with the quantum numbers
$I=1,0^{-+}$ and $I=0,0^{-+}$ are well seen. We remind that there are
no vacuum fermion loops in this large $N_c$ model and hence the $U(1)_A$
symmetry is broken only spontaneously. The spectrum exhibits 
approximate restoration
of the chiral symmetry in excited hadrons, for details we refer to \cite{WG2}
and for a review to ref. \cite{GPR}.
%
%\begin{widetext}
\begin{figure*}
\mbox{
\includegraphics[width=0.16\hsize]{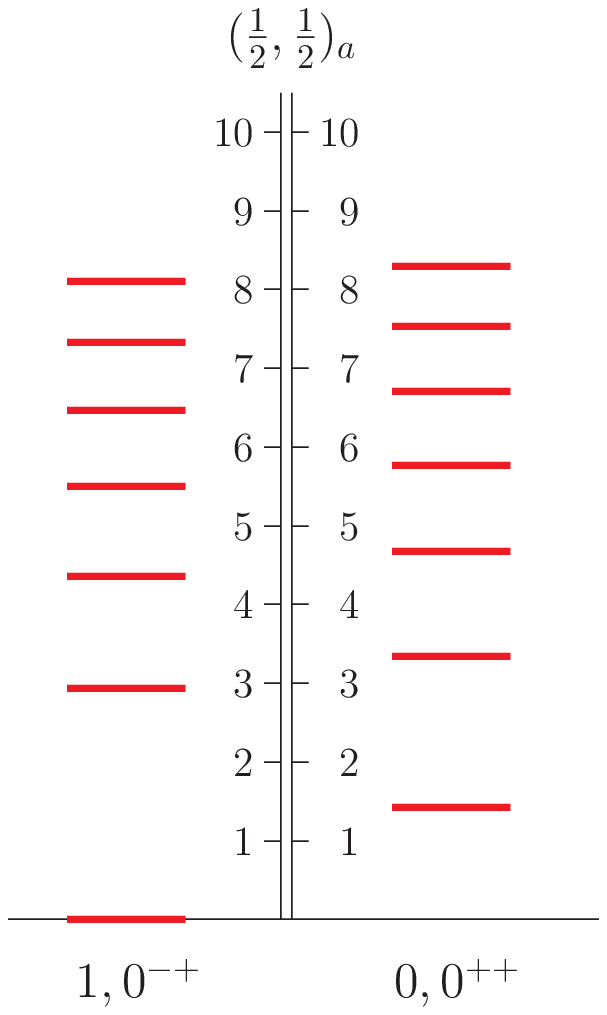}\,%
\includegraphics[width=0.16\hsize]{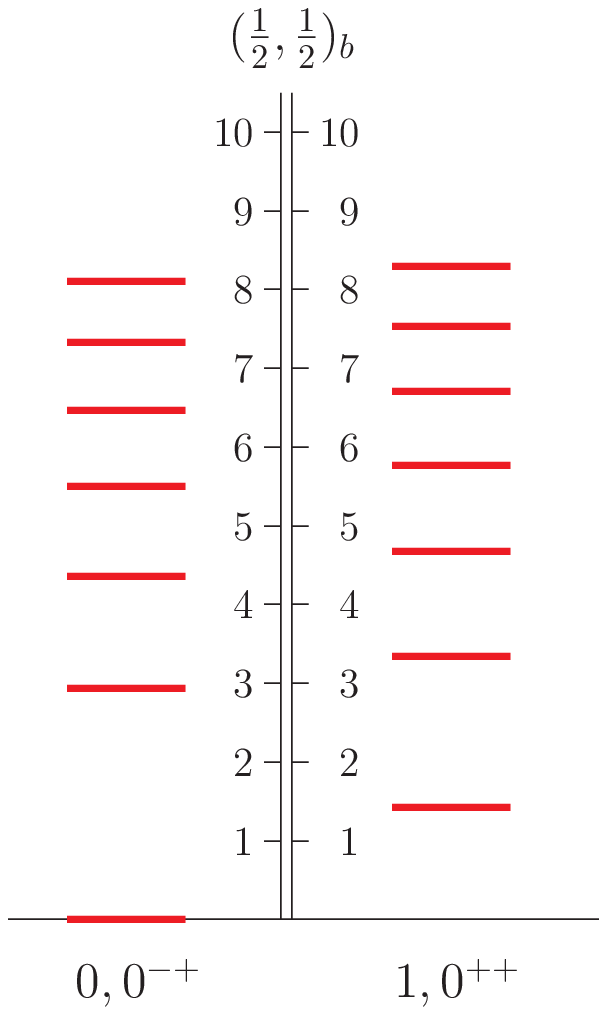}\,%
\includegraphics[width=0.16\hsize]{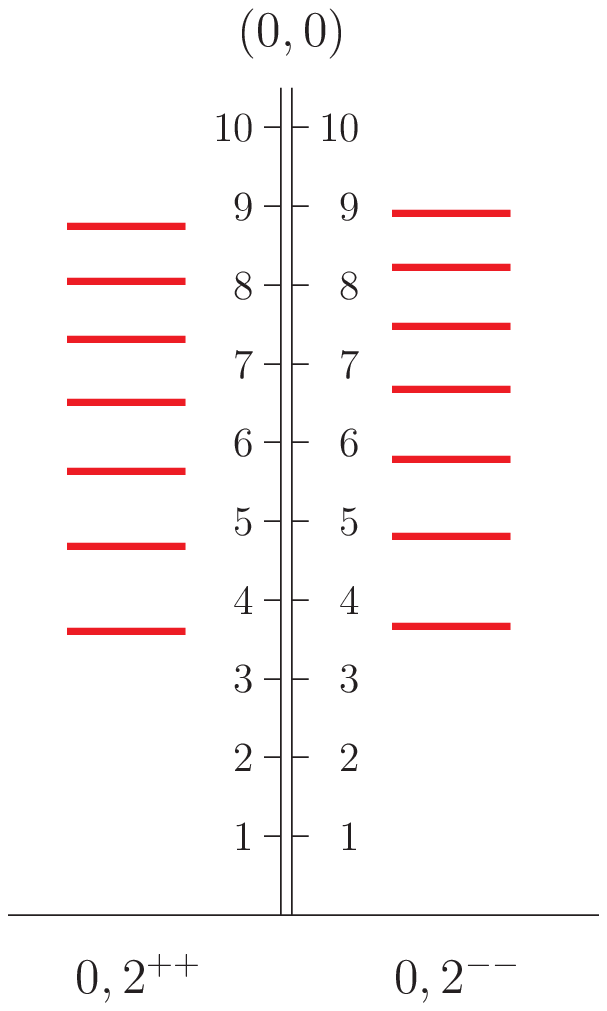}\,%
\includegraphics[width=0.16\hsize]{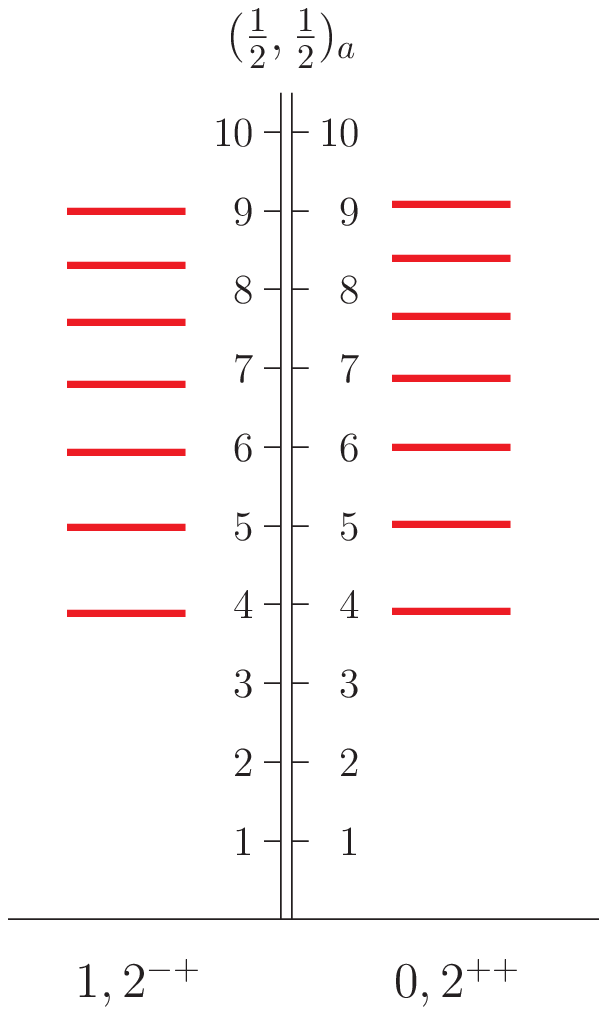}\,%
\includegraphics[width=0.16\hsize]{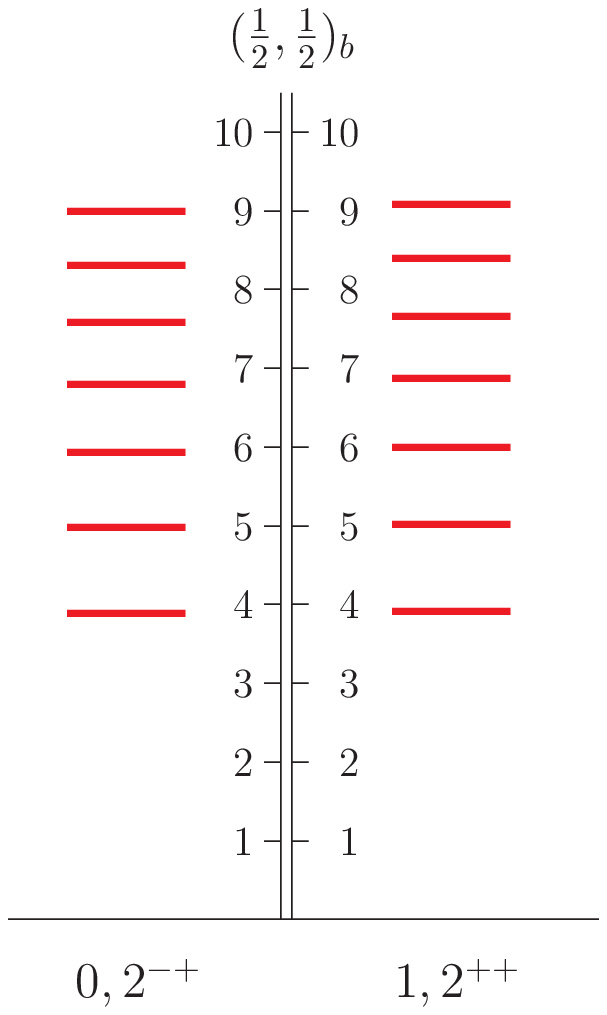}\,%
\includegraphics[width=0.16\hsize]{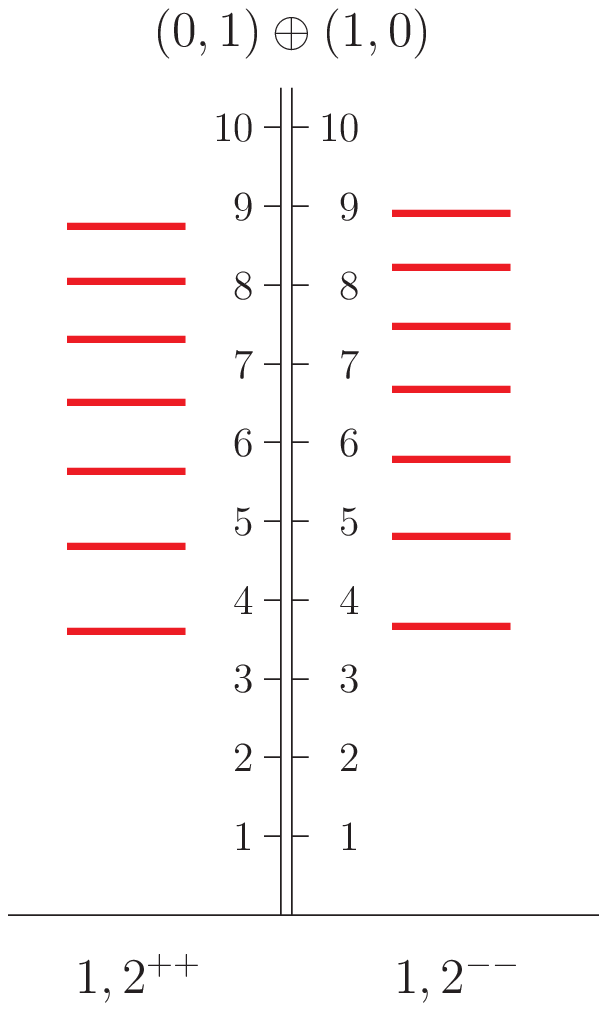}
}
\caption{Spectra for  $p_f=0.05\sqrt{\sigma}$. 
For $J=0$ only the two $(\frac{1}{2},\frac{1}{2})$ multiplets are 
present (left two panels). For $J>0$ there are also the 
$(0,0)$ and $(0,1)\oplus(1,0)$ multiplets. In the remaining four 
panels we show all multiplets for $J=2$. Masses are in units of $\sqrt \sigma$.
Meson quantum numbers are $I,J^{PC}$.}
\end{figure*}
\begin{figure*}
\mbox{
\includegraphics[width=0.16\hsize]{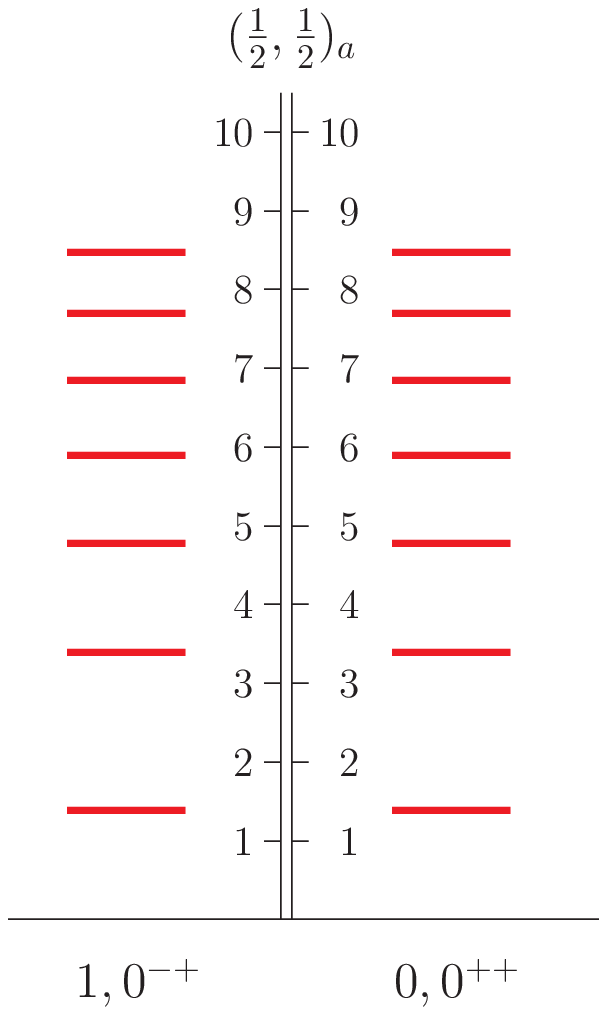}\,%
\includegraphics[width=0.16\hsize]{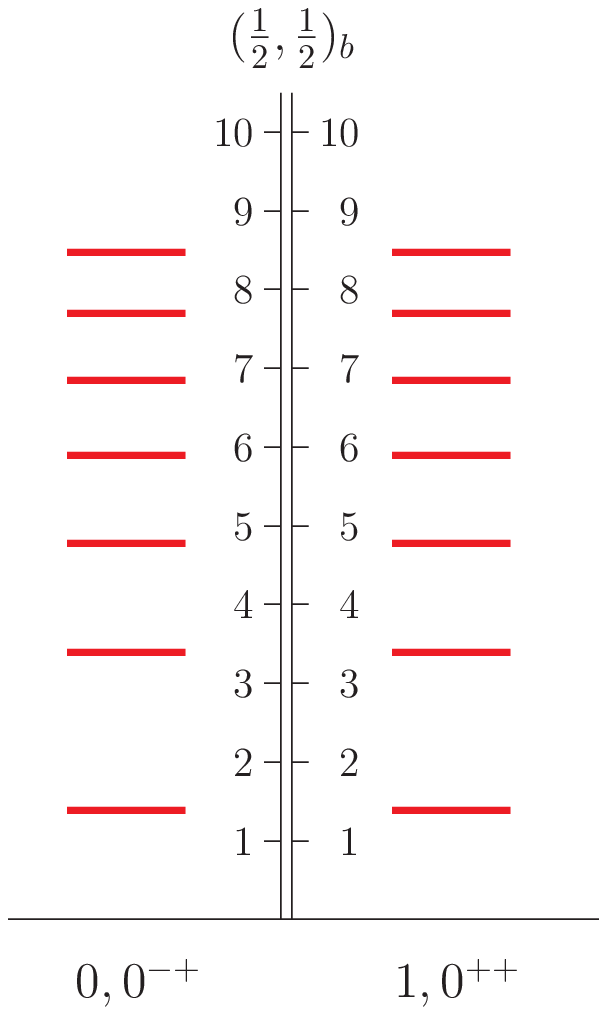}\,%
\includegraphics[width=0.16\hsize]{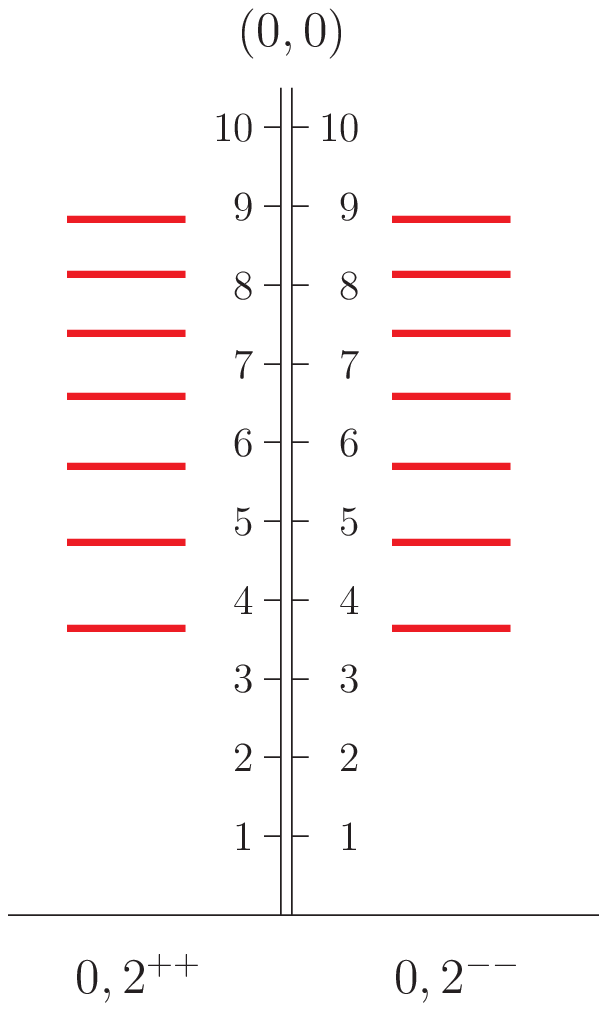}\,%
\includegraphics[width=0.16\hsize]{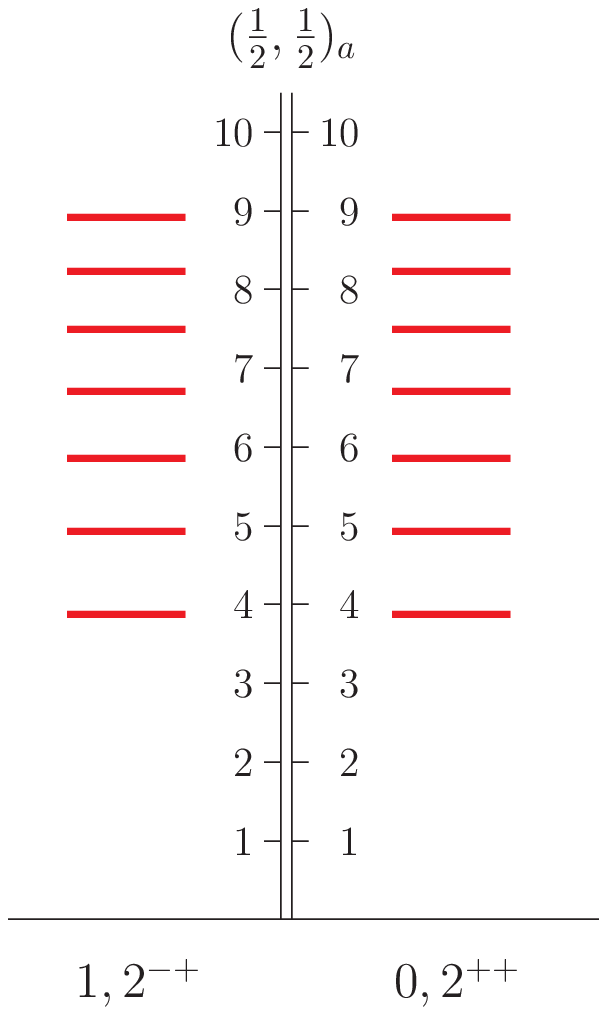}\,%
\includegraphics[width=0.16\hsize]{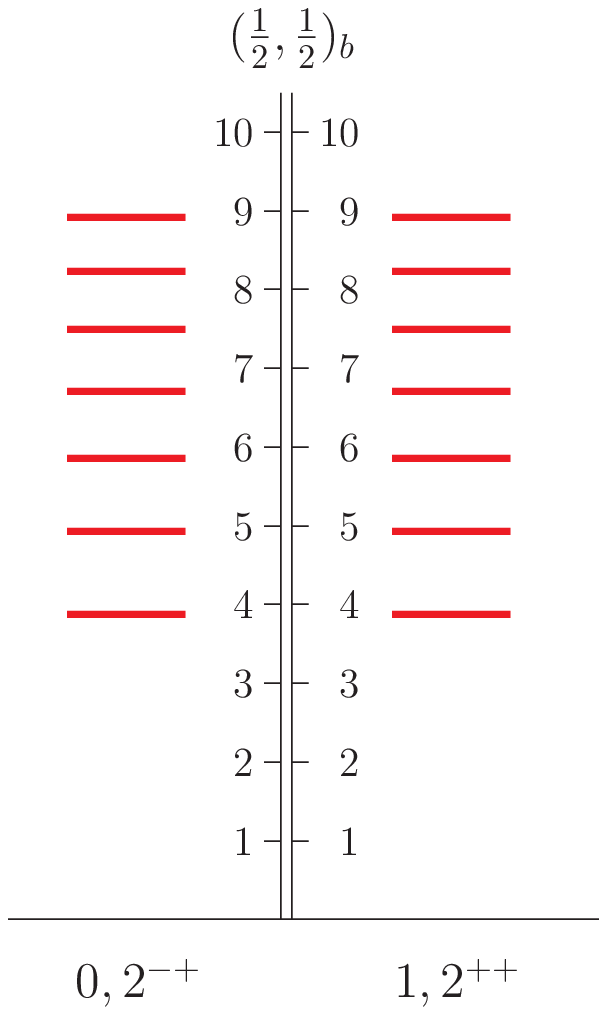}\,%
\includegraphics[width=0.16\hsize]{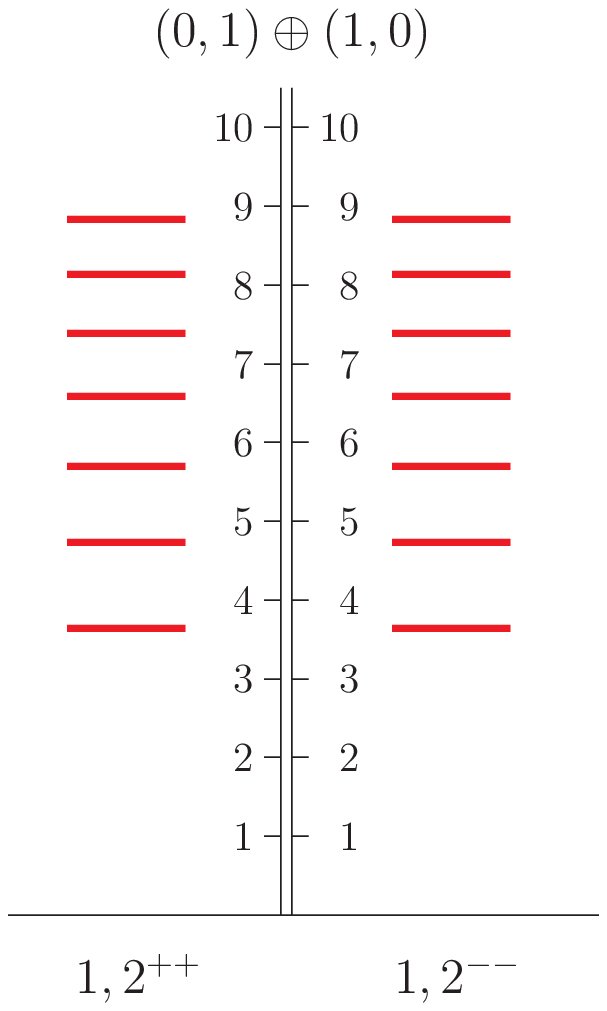}
}
\caption{As Fig. 4 but for Fermi momentum $p_f=0.2\sqrt{\sigma}$.}
\end{figure*}
%\end{widetext}
%

The spectrum above the critical Fermi momentum, i.e. for 
$p_f = 0.2\sqrt{\sigma}$, is shown
in Fig. 5. This spectrum is qualitatively different as compared to Fig. 4.
All the states are in {\it exact} chiral multiplets. 
A fundamental symmetry reason of this degeneracy is rather obvious:
When the vacuum is chiral-invariant, i.e., we are in the Wigner-Weyl
mode, the bound state spectrum (if it exists) can only be in exact
chiral multiplets \cite{CL}. In our case in the Wigner-Weyl mode the quarks
are confined and the bound states do exist. Consequently the resulting
color-singlet spectrum is manifestly chirally-symmetric.
This also directly follows from the analytical symmetry properties of the
Bethe-Salpeter equation in the chiral symmetry limit $M(p)=0$
 \cite{WG2}.

Though the chiral symmetry
is manifestly restored, one observes
{\it finite-energy well defined hadrons}. Obviously the mass generation
mechanism in these hadrons has nothing to do with the chiral symmetry breaking
and is not related with the quark condensate. The mass
generation mechanism for these chirally symmetric hadrons 
comes from the manifestly chirally symmetric dynamics and
is similar to
the high-lying states in the chiral symmetry broken phase on Fig. 4.  

On Fig. 6 we show an evolution of the ground state pion and the 
sigma mesons as a function of the Fermi momentum. 
\begin{figure}
\includegraphics[width=0.8\hsize,clip=]{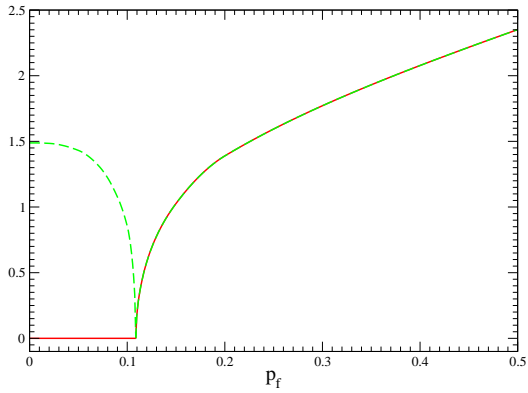}
\caption{Masses of the pseudoscalar (solid) and scalar (dashed) mesons 
in units of $\sqrt \sigma$ as functions of the Fermi momentum
in the same units.}
\end{figure}

\section{Conclusions}
 As a conclusion we have demonstrated
that it is possible to have at finite chemical potential a confining but 
chirally symmetric matter
consisting of chirally symmetric hadrons. For that we used the only
known exactly solvable model that is confining and manifestly chirally
symmetric. Certainly this model is not QCD and perhaps confinement
mechanism in QCD is more complicated. However, what we have shown, is
"a matter of principle demonstration". We have also clarified perhaps
a generic mechanism of the phenomenon. Namely, the chiral symmetry 
restoration
means that the Lorentz-scalar part of the quark self-energy vanishes.
But it does not require yet that the other parts of the quark self-energy
vanish either. If they do not and the quark is still off-shell, then
it cannot be observed and such a single quark is confined. At the
same time the color-singlet hadrons are well-defined and finite on-shell
systems. 

 Whether this happens 
in QCD or not is still an open question but it definitely suggests that 
there are no
reasons to believe that deconfinement and chiral restoration necessarily
coincide at finite density. If such a chirally symmetric but confining
phase does exist in QCD, then it will imply dramatic modifications
of the QCD phase diagram. Schematically this diagram would look like
on Fig. 7.

\begin{figure}
\includegraphics[width=0.8\hsize,clip=]{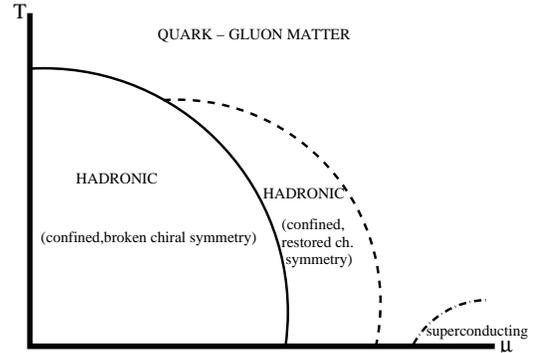}
\caption{Schematic possible phase diagram in the $T - \mu$ plane.}
\end{figure}

 It will also have
significant implications for astrophysics: The interactions of these
chirally-symmetric hadrons can be only of short-range as they decouple
from the Goldstone bosons and their  weak decay rate is quite different
since their axial  charge vanishes \cite{gl}.

\medskip
{\bf Acknowledgements}
We are thankful to Tom Cohen, Rob Pisarski and Kai Schwenzer
for discussions. L. Ya. G. acknowledges support of the Austrian Science
Fund through the grant P19168-N16.

\end{document}